\documentclass[a4paper, 14pt, conference,onecolumn]{ieeeconf}      

\IEEEoverridecommandlockouts                              

\usepackage{graphics} 
\usepackage{epsfig} 

\usepackage{amsmath} 
\usepackage{amssymb}  
\usepackage{tabularx}
\usepackage{amsmath}
\usepackage{balance}

\newcounter{subassumption}[asu]

\makeatletter
\renewcommand{\p@subassumption}{\theasu}
\makeatother

\usepackage{rotating,booktabs,multirow}
\makeatletter
\let\NAT@parse\undefined
\makeatother
\usepackage{hyperref}
\usepackage{graphicx}
\newtheorem{remark}{Remark}[section]

\usepackage[latin1]{inputenc}
\usepackage{color}
\usepackage{epsfig}
\usepackage{graphicx}
\usepackage[hang]{caption2}
\usepackage[normalsize,it,hang]{subfigure}
\DeclareGraphicsExtensions{.jpg,.png,.jpg,.jpg}

\title{\LARGE \bf
An alternative to proportional-integral and proportional-integral-derivative regulators: \\ Intelligent proportional-derivative regulators 
}

\author{Michel Fliess$^{1, 3}$, C\'{e}dric Join$^{2, 3}$
\thanks{$^{1}$LIX (CNRS, UMR 7161), \'Ecole polytechnique, 91128 Palaiseau, France. {\tt \small Michel.Fliess@polytechnique.edu } } 
\thanks{$^2$CRAN (CNRS, UMR 7039)), Universit\'{e} de Lorraine, BP 239, 54506 Vand{\oe}uvre-l\`{e}s-Nancy, France. \newline
{\tt\small Cedric.Join@univ-lorraine.fr}}
\thanks{$^{3}$AL.I.E.N. (ALg\`ebre pour Identification \& Estimation Num\'eriques), 7 rue Maurice Barr\`{e}s, 54330 V\'{e}zelise, France. \newline
        {\tt \small \{cedric.join, michel.fliess\}@alien-sas.com}}
        }

\begin{document}
\maketitle
\thispagestyle{empty}
\pagestyle{empty}

\begin{abstract}
This paper suggests to replace PIs and PIDs, which play a key r\^{o}le in control engineering, by intelligent Proportional-Derivative feedback loops, or iPDs, which are derived from model-free control. This standpoint is enhanced by a laboratory experiment.

\end{abstract}

\newpage

\section{Introduction}
Proportional-Integral-Derivative and Proportional-Integral regulators, or PIDs and PIs, are, as well-known, the most popular industrial feedback loops (see, \textit{e.g.}, \cite{astrom,murray,fol,franklin,odwyer,prouvost}). They ought therefore to be termed 
\emph{universal}. Their serious shortcomings partly explain nevertheless the important developments of ``modern'' model-based control theory during the second half of the twentieth and the beginning of the twenty first centuries. In spite of beautiful achievements, PIs and PIDs still largely prevail in engineering. This paper aims to suggest a new universal feedback loop. It is derived from \emph{Model-Free Control}, or \emph{MFC}, in the sense of \cite{csm}, \textit{i.e.}, a control setting with two purposes:
\begin{itemize}
\item to keep the benefits of PIs and PIDs, and, especially, the futility of almost any mathematical modeling in control engineering;
\item the deletion of many drawbacks of PIs and PIDs like, for instance, the devilish gain tuning difficulties or the lack of robustness.
\end{itemize}
The numerous successful concrete applications (see, \textit{e.g.}, \cite{bara,csm,ml}, and the corresponding bibliographies for references until the beginning of 2020) demonstrate that those aims have been fulfilled to a great extent. See, \textit{e.g.}, the recent appraisal in\cite{onr}: ``MFC is computationally efficient, easily deployable even on small embedded devices, and can be implemented in real time.'' Several conclusive concrete comparisons with PIs and PIDs have been published (see, \textit{e.g.}, \cite{agee1,agee2,ignition,mercer,brest,hong,mx,kizir,michel,ndoye,park,tide,huma,madrid}).\footnote{Comparisons with several other control strategies have also been investigated (see, \textit{e.g.}, \cite{abba,onr,barth,adrc,jama,toulon,wang,yang,zhang1,zhang2}).} Here we suggest a {\bf single} universal feedback loop, \textit{i.e.}, the \emph{intelligent Proportional-Derivative} controller, or \emph{iPD},
\begin{equation}\label{ipd}
u = - \frac{F_{\text{est}} - \ddot{y}^\ast + K_P e + K_D \dot{e}}{\alpha}
\end{equation}
which is derived from the \emph{ultra-local model} of order $2$ in the sense of \cite{csm}
\begin{equation}\label{2}
\ddot{y} = F + \alpha u
\end{equation}
\begin{itemize}
\item $u$ and $y$ are the input and output variables.
\item The time-varying quantity $F$ corresponds to the poorly known plant and to the disturbances; $F_{\text{est}}$ is an estimate of $F$.
\item The constant $\alpha$ is chosen such that the three terms of Equation \eqref{2} are of the same magnitude.
\item $y^\ast$ is the reference trajectory.
\item $e = y - y^\ast $ is the tracking error.
\item The constants $K_P$ and $K_D$ are the gains.
\end{itemize}
Successful implementation of iPDs have already been achieved a number of times (see, \textit{e.g.}, \cite{ignition,barth,haddar,sinofrench,menhour,sun,ferry}):
\begin{itemize}
\item The gain tuning is straightforward.
\item Recent algebraic parameter identification techniques\cite{garnier,sira} and Riachy's trick \cite{riachy} permit to avoid tedious numerical derivations of the output signal $y$.
\item Good robustness with respect to corrupting noises is ensured via a new noise understanding \cite{noise}.
\end{itemize}

Until today MFC was offering not only the iPD \eqref{2} but also the \emph{intelligent Proportional} controller, or (\emph{iP}),
\begin{equation}\label{ip}
u = - \frac{F_{\text{est}} - \dot{y}^\ast + K_P e}{\alpha}
\end{equation}
for the \emph{ultra-local model} of order $1$\cite{csm},
\begin{equation}\label{un}
\dot{y} = F + \alpha u
\end{equation}
Assume that ``almost'' any I/O system may be locally approximated by Equation \eqref{2}, \textit{i.e.}, by an ultra-local model of order $2$. This postulate, which is the main conceptual novelty of our paper, is vindicated by fifteen years of MFC practice. It applies of course to a system which is well approximated by an ultra-local-model of order $1$, \textit{i.e.}, by Equation \eqref{un}. 
The ubiquity of the iPD \eqref{ipd} follows at once. A laboratory experiment is presented in order to sustain this viewpoint. The Quanser AERO, \textit{i.e.}, a half-quadrotor, is used as in two recent publications\cite{ml,med}: it is available at the \textit{Universit\'{e} de Lorraine}. The results shows that adding an integrator deteriorates the behavior: intelligent Proportional-Integral-Derivative controllers, or iPIDs, might be useless.

Our paper is organized as follows. Basic facts on MFC are reviewed in Section \ref{basic}. Riachy's trick \cite{riachy} in particular, which plays a key r\^{o}le in the implementation of iPDs, is detailed in Section \ref{riac} for the first time since a conference communication ten years ago where it was only sketched. Section \ref{lab} is dedicated to the laboratory experiments. Concluding remarks and suggestions for future investigations are presented in Section \ref{conclusion}.



\section{Basic facts on model-free control}\label{basic}
\subsection{Generalities}
Elementary functional analysis and differential algebra permit to prove\cite{csm} that under quite weak assumptions any SISO system with input $u$ and output $y$ may be well approximated by
\begin{equation}\label{siso}
y^{(n)} = F + \alpha u
\end{equation}
where
\begin{itemize}
\item the system might be infinite-dimensional and correspond to rather arbitrary functional equations and/or partial differential equations;
\item $n \geq 1$ is the derivation order;
\item the time-varying quantity $F$ subsumes not only the un-modeled dynamics, but also the external disturbances;
\item the constant $\alpha \in {\mathbb R}$ is such that the three quantities $y^{(n)}$, $F$, $\alpha u$ in  Equation \eqref{siso} are of the same order of magnitude.
\end{itemize} 
Note that 
\begin{itemize}
\item the poorly known plant is  not necessarily of order $n$: $y^{(\nu)}$, where $\nu > n$ may be sitting in $F$;
\item in all the numerous concrete case-studies that were encountered until now, $n = 1$ or $2$;
 \item it is meaningless to try to estimate $\alpha$ precisely.
\end{itemize}
If $n = 2$, Equation \eqref{siso} yields Equation \eqref{2}. 
Associate to Equation \eqref{2} the \emph{intelligent Proportional-Integral-Derivative} controller, or \emph{iPID},
\begin{equation}\label{ipid}
u = - \frac{F_{\text{est}} - \ddot{y}^\ast + K_P e + K_I \int e + K_D \dot{e}}{\alpha}
\end{equation}
where 
\begin{itemize}
\item $y^\ast$ is the reference trajectory,
\item $e = y - y^\ast$ is the tracking error,
\item $K_P, K_I, K_D$ are tuning gains.
\end{itemize}
Equations \eqref{2} and \eqref{ipid} yield
\begin{equation}\label{track2}
\ddot{e} + K_D \dot{e} + K_P e + K_I \int e = F - F_{\text{est}} 
\end{equation}
Select $K_P$, $K_I$, $K_D$ such that the roots of the characteristic equation
\begin{equation}\label{car}
s^3 + K_D s^2 + K_P s  + K_I = 0
\end{equation}
have strictly negative real parts. It ensures that $\lim_{t \to +\infty} e(t) \approx 0$ if the estimate $F_{\text est}$ is ``good,'' \textit{i.e.}, $F - F_{\text{est}} \approx 0$. Thus \emph{local stability} around the reference trajectory is trivially ensured via this feedback loop. Note that \emph{global stability} is obviously difficult, if not impossible, to study without any mathematical modeling. Let us emphasize that the situation is much worse with classic PIs and PIDs where no stability property is available in general. The \emph{intelligent Proportional-Derivative} controller (\emph{iPD})  \eqref{ipd}) is obtained by setting $K_I = 0$. The \emph{intelligent Proportional-Integral} (iPI) (resp. \emph{intelligent Proportional} (\emph{iP})) controller corresponds to $K_D = 0$ (resp. $K_I = K_D = 0$).

The application of the Routh-Hurwitz  \cite{franklin,gantmacher,prouvost} criterion to Equation \eqref{car} shows that the tuning of $K_P$ and $K_I$ alone does not permit to get arbitrary roots. It yields\\\\
\textbf{Fact 1.} If $n = 2$, iPs and iPIs do not work in general.\\

\subsection{An estimation integral}
Take Equation \eqref{2}. According to a a classic result of mathematical analysis (see, \textit{e.g.}, \cite{bourbaki}), $F$ may be approximated, under a weak integrability condition, by a piecewise constant function. Replace therefore Equation \eqref{2} by  
\begin{equation*}\label{2bis}
\ddot{y} = \Phi + \alpha u
\end{equation*}
where $\Phi$ is a constant. It yields via the rules of operational calculus (see, \textit{e.g.}, \cite{yosida}) 
$$
s^2 Y = \frac{\Phi}{s} + U + sy(0) + \dot{y}(0)
$$
In order to get rid of the initial conditions, derive both sides twice by $\frac{d}{ds}$:
$$
\frac{2\Phi}{s^3} = s^2 \frac{d^2 Y}{ds^2} + 4s \frac{dY}{ds} + 2Y - \frac{d^2 U}{ds^2}  
$$
Multiplying both sides by $s^{-N}$, $N \geq 3$, permits to get rid of positive powers of $s$, \textit{i.e.}, of time derivatives, which are very sensitive to corrupting noises. Negative powers of $s$, which correspond to iterated time integrals: noises are mitigated (see Remark \ref{robust} below). Thanks to the correspondance between $\frac{d}{ds}$ and the multiplication by $- t$, we get in the time domain for $N = 3$ (see, \textit{e.g.}, \cite{barth,menhour})
\begin{equation}\label{int2}
{F_{\text{est}}(t)  = \frac{60}{\tau^5}} \int_{t - \tau}^{t} (\tau^2 + 6\sigma^2 - 6\tau\sigma)y(\sigma)d\sigma  - \frac{30\alpha}{\tau^5} \int_{t - \tau}^{t}  (\tau - \sigma)^2 \sigma^2 u(\sigma) d\sigma
\end{equation}
where $\tau > 0$ is ``small.''
\begin{remark}\label{robust}
According to \cite{noise} the robustness with respect to corrupting noises is ensured via the integrals in Equation \eqref{int2}. There is no need to know the precise probabilistic/statistical nature of the noises which are viewed as quick fluctuations around $0$. See, \textit{e.g.}, \cite{morales,sira} for applications in signal processing. 
\end{remark}
\begin{remark}
In practice the integrals in Equation \eqref{int2} are replaced by \emph{finite impulse response} (\emph{FIR}) filters (see, \textit{e.g.}, \cite{rabiner}).
\end{remark}

\subsection{iPs \& PIs, iPD \& PIDs}\label{connection}
\subsubsection{PI \& iP}\label{1}
Consider the PI regulator
$$
u = k_p e + k_i \int e
$$
where $k_p$, $k_i$ are the gains. Derive both sides:
$$
\dot{u}(t) = k_p \dot{e}(t) + k_i e(t)
$$
A crude sampling yields:
\begin{equation}\label{sampling1}
\frac{u(t) - u(t - h)}{h} = k_p \frac{e(t) - e(t - h)}{h} + k_i e(t)
\end{equation}
where $h$ is the sampling period.



Replace $F$ in Equation \eqref{ip} by ${\dot y}(t) - \alpha u (t-h)$ and therefore by
$$\frac{y(t) - y(t-h)}{h} - \alpha u (t-h)$$
It yields
\begin{equation}
\label{eqDiscr_i-POne} u (t) = u (t - h) - \frac{e(t) -
e(t-h)}{h\alpha} + \dfrac{K_P}{\alpha}\, e(t)
\end{equation}

\textbf{Fact 2.}
Equations \eqref{sampling1} and
\eqref{eqDiscr_i-POne} become identical if we set
\begin{align}
\label{eqPI_i-P_corresp} k_p &= - \dfrac{1}{\alpha h}, \quad k_i =
\dfrac{K_P}{\alpha h}
\end{align}

\subsubsection{PID \& iPD}
Consider the PID regulator
$$
u = k_p e + k_i \int e + k_d \dot{e}
$$
where $k_p$, $k_i$, $k_d$ are the gains. Derive both sides:
$\dot u(t) = k_p {\dot e}(t) + k_i e(t) + k_d {\ddot e}(t)$. It
yields the obvious sampling
\begin{equation}\label{sampling2}
\frac{u(t) - u(t - h)}{h} = k_p \dot{e}(t) + k_i e(t) + k_d \ddot{e}(t)
\end{equation}

Replace in Equation \eqref{2} $F$ by
$\ddot{y}(t) - \alpha u(t - h)$:
\begin{equation}\label{iPDd}
\frac{u (t) - u (t - h)}{h} = - \dfrac{1}{\alpha h} {\ddot e}(t)  +
\dfrac{K_P}{\alpha h} e(t) + \dfrac{K_D}{\alpha h} {\dot e}(t)
\end{equation}

\textbf{Fact 3.}
Equations \eqref{sampling2} and \eqref{iPDd}
become identical if we set
\begin{equation}
\label{eqPID_i-PD_corresp} k_p = \dfrac{K_D}{\alpha h}, \quad k_i =
\dfrac{K_P}{\alpha h}, \quad k_d = - \dfrac{1}{\alpha h}
\end{equation}

\subsubsection{Sampling and equivalence}
The equivalences in Facts 2 and 3, which go back to \cite{andrea,csm}, are strictly related to
time sampling, {\it i.e.}, to the computer implementation, as
demonstrated by taking $h \downarrow 0$ in Equations
\eqref{eqPI_i-P_corresp} and \eqref{eqPID_i-PD_corresp}. In other words, the above equivalences do not hold in continuous-time. Those two facts nevertheless
\begin{enumerate}
\item explain the ubiquity in ``real life'' of PIs and PIDs,
\item explain the difficulty of gain tuning for PIs and PIDs,
\item question the necessity of introducing generalized PIDs (see, \textit{e.g.}, \cite{huba,frac}).
\end{enumerate}

\begin{remark}
Formulae \eqref{eqPI_i-P_corresp} and \eqref{eqPID_i-PD_corresp} should not be employed (compare with \cite{stankovic2}) in order to deduce an iP (resp. a PI) from a PI (resp. an iP).
\end{remark}

\subsection{Riachy's trick}\label{riac}
Consider the ultra-local model \eqref{2} and the iPID \eqref{ipid}. Following Riachy \textit{et al.}\cite{riachy} write
\begin{equation}\label{trick}
\ddot{y} + K_D \dot{y} = F +  K_D \dot{y} + \alpha u
\end{equation}
Set 
\begin{equation}\label{c}
Y(t) = y(t) + K_D \int_{c}^{t} y(\sigma) d\sigma 
\end{equation}
where $0 \leq c < t$. It yields
$$
\ddot{Y} = \ddot{y} + K_D \dot{y} 
$$
Set
$$
{\frak F} = F + K_D \dot{y} 
$$
Equations \eqref{2}-\eqref{trick} read
\begin{equation}\label{riachy}
\ddot{Y} = {\frak F} + \alpha u
\end{equation}
The iPID \eqref{ipid} becomes
\begin{equation}\label{ipidbis}
u = - \frac{{\frak F}_{\text est} - \ddot{y}^\ast + K_P e + K_I \int_c^t e - K_D \dot{y}^\ast}{\alpha}
\end{equation}
${\frak F}_{\text{est}}$ is given by \eqref{int2} where $y$ is replaced by $Y$. Set $K_I = 0$ in Equation \eqref{ipidbis} for the iPD. 
Riachy's trick permits to implement iPDs and iPIDs without estimating derivatives. 
\begin{remark}
Resetting $c$ in Equation \eqref{c} permits to trivially avoid any integral windup in Equations \eqref{c} and \eqref{ipidbis}.
\end{remark}

\subsection{The key postulate}
Assume that almost any system encountered in practice, which is well approximated by Equation \eqref{siso}, where $n \geq 1$, may be also well approximated by Equation \eqref{2}, \textit{i.e.}, by an ultra-local model of order $2$.
\begin{remark}
The determination of the lowest order $n$ in Equation \eqref{siso} seems to be a difficult mathematical question\cite{n=2} which is far from being fully clarified. The above postulate shows that this question may be pointless from an applied standpoint. 
\end{remark}

\section{A laboratory experiment}\label{lab}
\subsection{Presentation}
\begin{figure*}
\center
\includegraphics[width=0.5\textwidth]{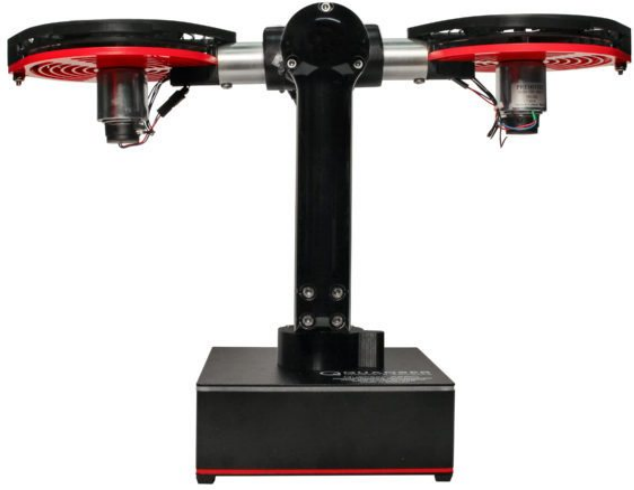}
\caption{The Quanser AERO} \label{maquette}
\end{figure*}
Employ as in \cite{ml,med} the \textit{Quanser AERO} (see Figure \ref{maquette}), \textit{i.e.}, a half-quadrotor, manufactured by the company Quanser  (see the link {\tt https://www.quanser.com/products/quanser-aero/}).
Two motors driving the propellers, which might turn clockwise or not, are controlling the angular position $y$ (rad) of the arms. 

Write $v_i$, $i = 1, 2$, the supply voltage of motor $i$, where $- 24{\rm v} \leq v_i \leq 24{\rm v}$ (volt). Introduce the single control variable $u$. Set
\begin{itemize}
\item if $u>0$, then $v_1=10+u$ and $v_2=-10-u$
\item if $u<0$, then $v_1=-10+u$ et $v_2=10-u$
\end{itemize}
The sampling time interval is $10$ms. The sample number for estimating integrals is $30$.

After exhibiting as in \cite{med} the performances of an iP, corresponding to a first order ultra-local model, we show that there is no degradation with an iPD associated to a second order ultra-local model. Let us emphasize that an iPID, which leads to oscillations, should be avoided. 
\begin{remark}
The experiments show, as outlined in Remark \ref{robust}, a good robustness with respect to the unavoidable corrupting noises, among which the quantization noise is perhaps the most important one.
\end{remark}




\subsection{iP}\label{ipscen}
\begin{figure*}[!ht]
\centering%
\subfigure[\footnotesize Reference trajectory and position ]
{\epsfig{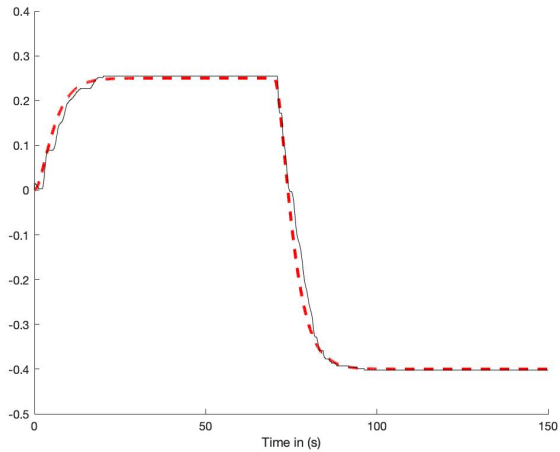}}
\subfigure[\footnotesize Control input ]
{\epsfig{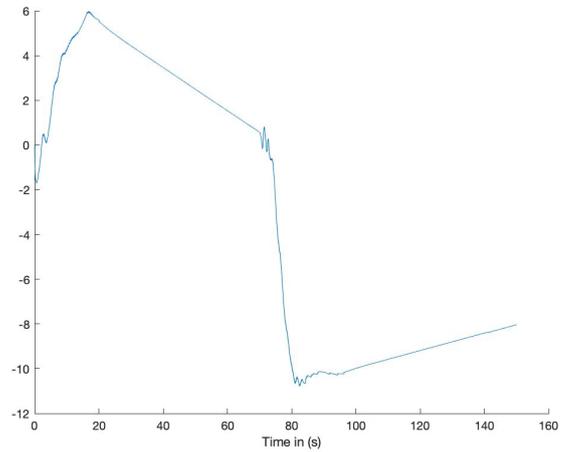}}
\caption{iP: Scenario 1}\label{S1}
\end{figure*}

\begin{figure*}[!ht]
\centering%
\subfigure[\footnotesize Reference trajectory and position ]
{\epsfig{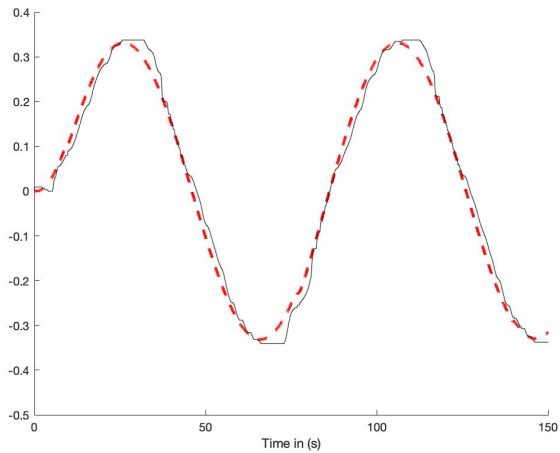}}
\subfigure[\footnotesize Control input ]
{\epsfig{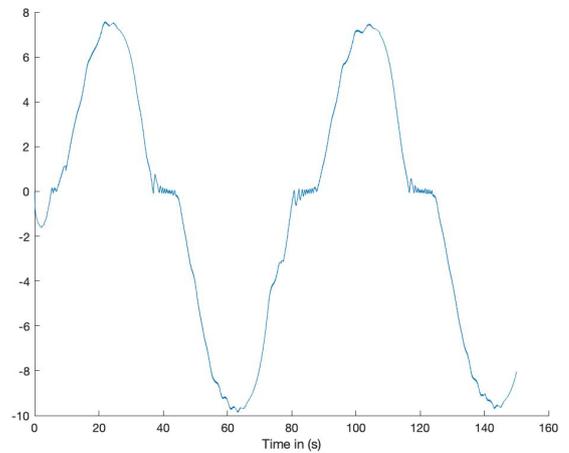}}
\caption{iP: Scenario 2}\label{S2}
\end{figure*}

\begin{figure*}[!ht]
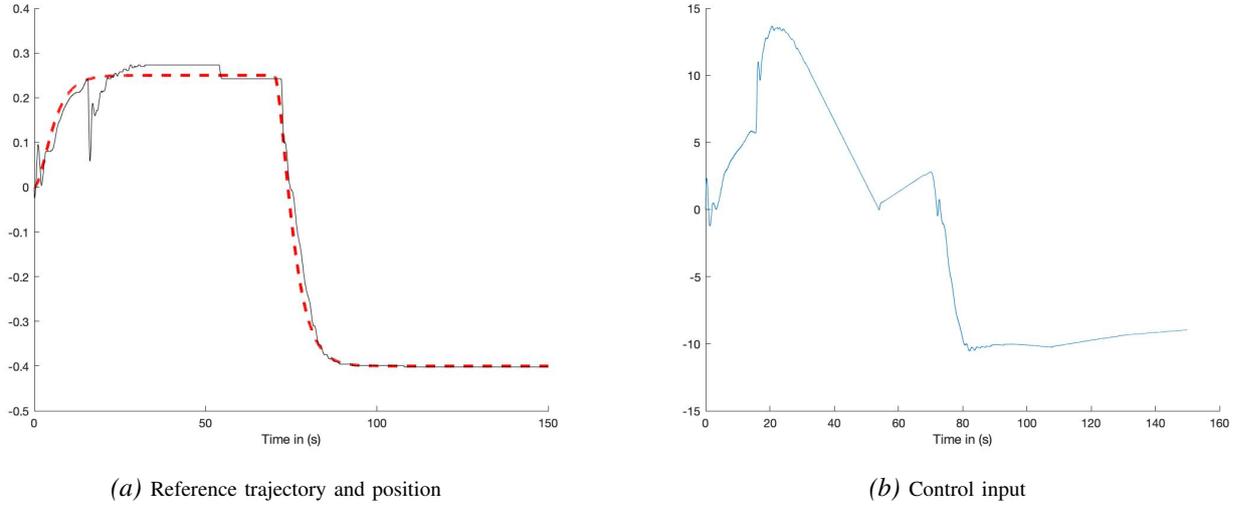

\centering%
\subfigure[\footnotesize Reference trajectory and position ]
{\epsfig{figure=S7Y.jpg,width=0.49\textwidth}}
\subfigure[\footnotesize Control input ]
{\epsfig{figure=S7C.jpg,width=0.49\textwidth}}
\caption{iP: Scenario 3}\label{S3}
\end{figure*}

Set in Equations \eqref{ip}-\eqref{un} $\alpha = 10$, $K_P = 25$. Three scenarios are examined:
\begin{itemize}
\item \textbf{Scenario 1:} Figure \ref{S1} exhibits good performances with a rather simple reference trajectory.
\item \textbf{Scenario 2:} With a more complex reference trajectory Figure \ref{S2} still displays good performances.
\item \textbf{Scenario 3:} At $t = 15$s, disrupt now manually the system so that it rotates around the base axis. The corresponding centrifugal force should be seen as a perturbation. Figure \ref{S3} shows an excellent rejection.
\end{itemize}




\subsection{iPD}\label{ipdscen}
\begin{figure*}[!h]
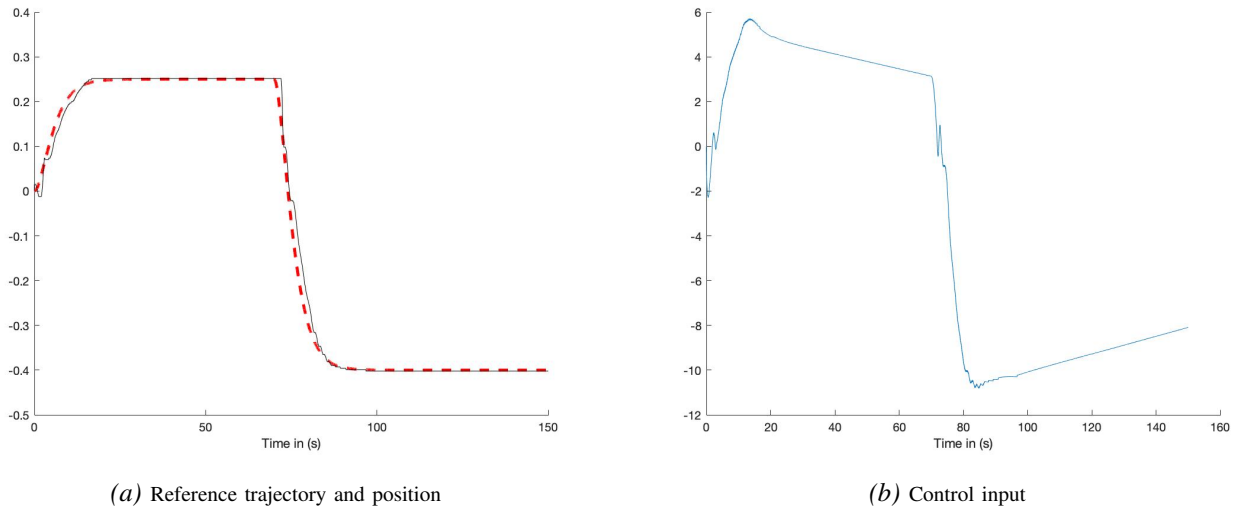

\centering%
\subfigure[\footnotesize Reference trajectory and position ]
{\epsfig{figure=S1Y.jpg,width=0.49\textwidth}}
\subfigure[\footnotesize Control input ]
{\epsfig{figure=S1C.jpg,width=0.49\textwidth}}
\caption{iPD: Scenario 4}\label{S4}
\end{figure*}

\begin{figure*}[!h]
\centering%
\subfigure[\footnotesize Reference trajectory and position ]
{\epsfig{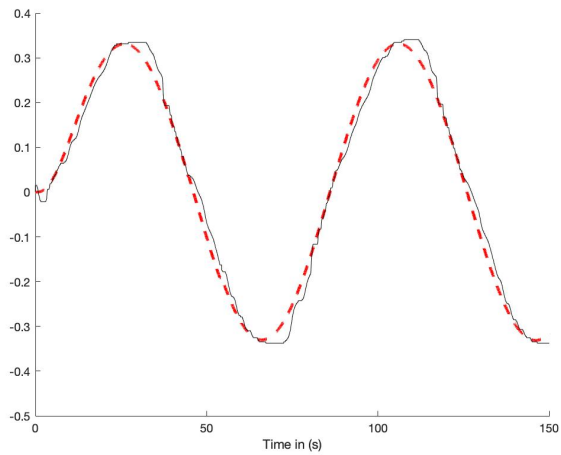}}
\subfigure[\footnotesize Control input ]
{\epsfig{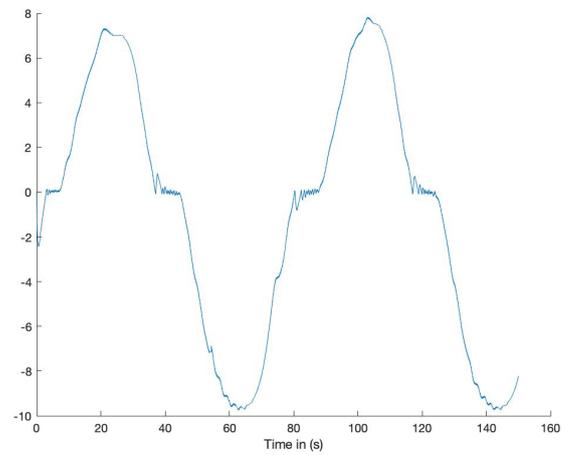}}
\caption{iPD: Scenario 5}\label{S5}
\end{figure*}

\begin{figure*}[!h]
\centering%
\subfigure[\footnotesize Reference trajectory and position ]
{\epsfig{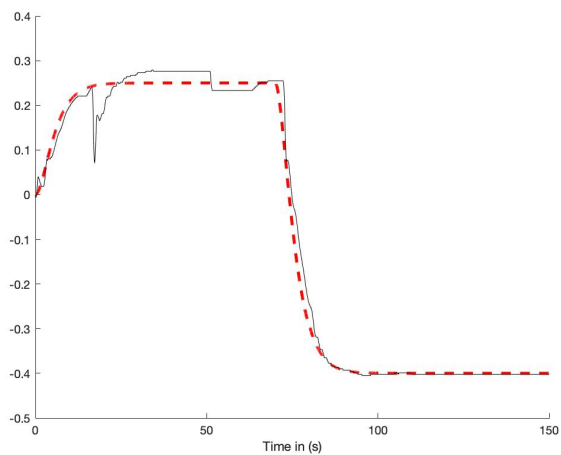}}
\subfigure[\footnotesize Control input ]
{\epsfig{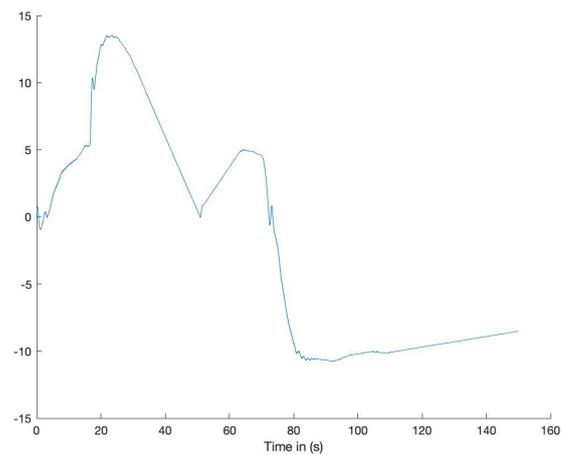}}
\caption{iPD: Scenario 6}\label{S6}
\end{figure*}

Employ now with the same device a second order ultra-local model and an iPD. Set $\alpha = 10$, $K_P = 25$, $K_D = 10$ in Equations \eqref{ipd}-\eqref{2}. Figures \ref{S4}, \ref{S5}, \ref{S6} display the performances with respect to scenarios $4$, $5$, and $6$, which are the analogues of scenarios $1$, $2$, and $3$ in Section \ref{ipscen}. The results are similar to those in Section \ref{ipscen}: they are again excellent.

\subsection{iPID}
\begin{figure*}[!h]
\centering%
\subfigure[\footnotesize Reference trajectory and position ]
{\epsfig{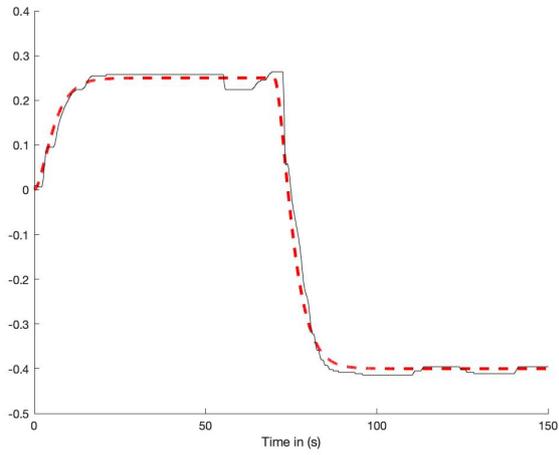}}
\subfigure[\footnotesize Control input ]
{\epsfig{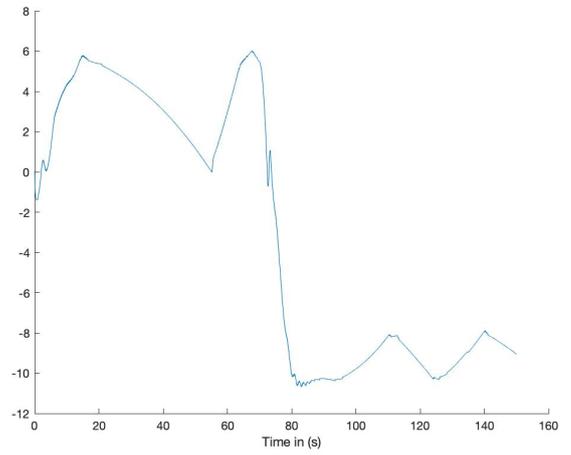}}
\caption{iPID: Scenario 7 ($K_I = 0.001$)}\label{S7}
\end{figure*}

\begin{figure*}[!h]
\centering%
\subfigure[\footnotesize Reference trajectory and position ]
{\epsfig{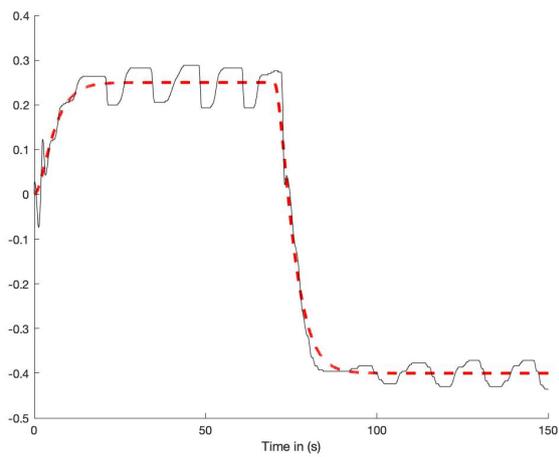}}
\subfigure[\footnotesize Control input ]
{\epsfig{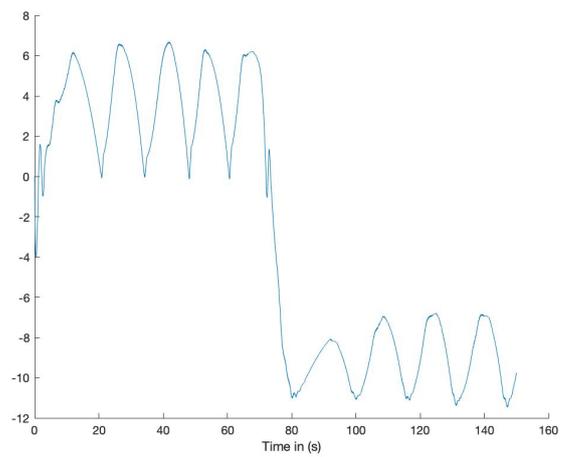}}
\caption{iPID: Scenario 8 ( $K_I = 0.01$)}\label{S8}
\end{figure*}

\begin{figure*}[!h]
\centering%
\subfigure[\footnotesize Reference trajectory and position ]
{\epsfig{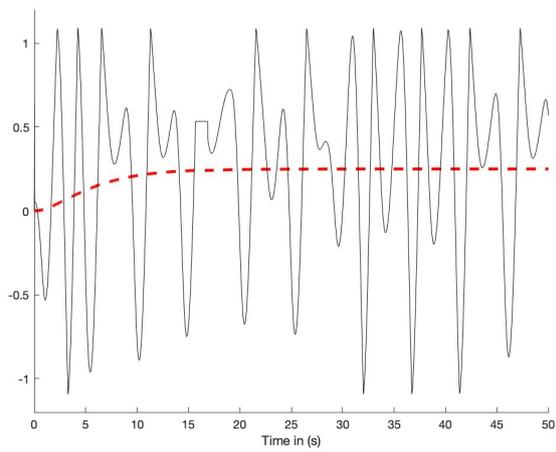}}
\subfigure[\footnotesize Control input ]
{\epsfig{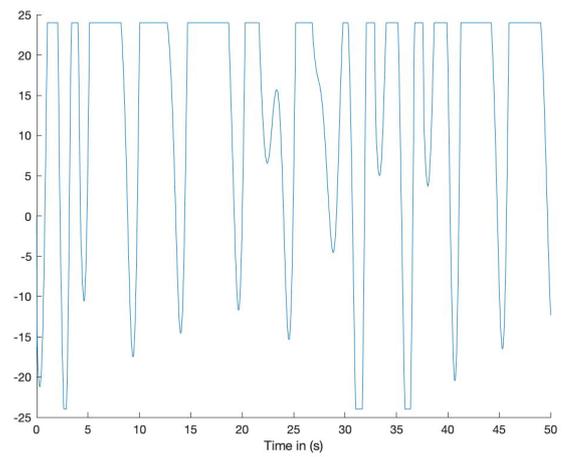}}
\caption{iPID: Scenario 9 ( $K_I = 0.1$)}\label{S9}
\end{figure*}

With respect to the iPD \eqref{ipid}, the scenarios $7$, $8$, and $9$, which are displayed by Figures \ref{S7}, \ref{S8}, and \ref{S9}, are the analogues of scenarios $1$ and $4$ in Sections \ref{ipscen} and \ref{ipdscen}. The coefficient $\alpha = 10$, the gains $K_P  = 25$ and $K_D = 10$ are the same as in Section  \ref{ipdscen}. Performances deteriorate when the gain $K_I$ increases ``too much.''  




\section{Conclusion}\label{conclusion}
In order to be fully accepted  the proposed replacement of traditional PIs and PIDs by iPDs needs of course further confirmations via many concrete case-studies. If this is the case, several investigations ought to be carried on. Let us mention here:
\begin{itemize}
\item The determination of the coefficient $\alpha$ in Equation \eqref{2} should be better analyzed. See \cite{ml} for a first hint. 
\item The impossibility of ensuring global stability in the context of model-free control should be replaced by a suitable replanning of the reference trajectory (see, \textit{e.g.}, \cite{replan} for a replanning example in the context of flatness-based control).
\item The extension to iPDs of the hardware for iPs in \cite{hardware}.
\item The extension of the treatment of multi-input multi-output (MIMO) systems in \cite{toulon}.
\item The extension to the discrete-time approach in \cite{sanyal}.
\item A more thorough study of iPIDs in order to determine possible benefits with respect to iPDs.\footnote{Would for instance the most convincing application \cite{jeong} to an electromagnetic robot, where an iPID is employed, be improved or not by an iPD?}
\item The regulation via PIDs of systems with delays is far from being fully understood (see, \textit{e.g.}, \cite{ma,silva}). It would be rewarding to extend, if possible, the time series setting in \cite{hamiche} for supply chain management.
\item The approach to machine learning sketched in \cite{ml} should benefit from the unicity of the feedback loop advocated here. See also \cite{sanyal}.
\end{itemize}

\end{document}